\documentclass{iau379}

\usepackage{amsmath}
\usepackage{graphicx}
\usepackage{multirow}

\begin{document}

\lefttitle{Bhattacharya et al.}
\righttitle{Weighing Andromeda: Mass estimates of the M~31 galaxy}

\jnlPage{1}{7}
\jnlDoiYr{2023}
\doival{10.1017/xxxxx}

\aopheadtitle{Proceedings of IAU Symposium 379}
\editors{P. Bonifacio,  M.-R. Cioni \& F. Hammer, eds.}

\title{Weighing Andromeda: Mass estimates of the M~31 galaxy}

\author{Souradeep Bhattacharya$^1$}
\affiliation{$^1$ Inter University Centre for Astronomy and Astrophysics, Ganeshkhind, Post Bag 4, Pune 411007, India}

\begin{abstract}
Andromeda (M~31) is the nearest giant spiral galaxy to our Milky Way,  and, over the past few decades, has been dubbed the most massive member of the Local Group. I explore the evolution of the measured mass of M~31 over the past $\sim80$~years, reviewing the different observational and modelling techniques that have developed over time to measure its mass. I discuss the best present-day constraints of the mass of M~31 and the consistency of different techniques. 
\end{abstract}


\maketitle

\section{Introduction}

The Andromeda galaxy (M~31) is the nearest giant spiral galaxy to our Milky Way (MW) and has long been considered its sister galaxy. Together they account for $\sim90\%$ of the mass of our Local Group (LG; \citealt{Pen16}). Given its large size and proximity to the MW, M~31 has long been the subject of many path-breaking studies in astronomy, the earliest of which was the first identification of ``island universes'' beyond the MW \citep{Hubble25,Hubble29}. Measured from Cepheid variables to be at a distance of 285~kpc \citep{Hubble25}, M~31 (along with the Triangulum galaxy, M~33) was the first of many galaxies to be studied, effectively kickstarting the field of Extragalactic Astrophysics. 

M~31 was also one of the first galaxies whose mass determination was attempted. \citet{Babcock39} constructed the rotation curve (RC) of M~31 out to $\sim100'$ from its center by measuring the radial velocities of its HII regions from spectroscopic observations. Assuming M~31 to be at a distance of 210~kpc\footnote{\citet{Babcock39} attributes this assumed distance to \citet{Hubble29}. However, the distance for M~31 quoted by \citet{Hubble29} is 275~kpc.} and modelling the disc of M~31 as rotating concentric flattened spheroids, \citet{Babcock39} measured the mass of M~31 as $10.2 \times 10^{10}$~M$_{\odot}$. \citet{Babcock39} also noted that the nearly constant angular velocity measured at the outer parts of M~31 was opposite to ``planetary'' type rotation, a problem that would later come to be known as the ``missing mass'' problem and nearly 30 years later would start the development of $\Lambda$CDM cosmology \citep{OP73}. 

In this paper, we will go through a near-exhaustive list of mass measurements of M~31. A number of galaxy mass measurement techniques, many of which find their first application to M~31, will be discussed in Section~\ref{sect:mass}. We will then compare the mass estimate of M~31 from different techniques, discuss possible reasons of discrepancies if any, and finally obtain the present-day best mass estimate of M~31 in Section~\ref{sect:disc}.

\section{Mass measurements of the Andromeda galaxy}
\label{sect:mass}

The various mass measurements of M~31 discussed in this paper have been noted in Table~\ref{table:mass} along with the original source of the measurement, the technique employed, the distance to M~31 assumed by the original source as well as the enclosed radius within which the mass has been measured (r$\rm_{enc}$; calculated assuming the present-day best known distance to M~31 of 776~kpc; \citealt{Savino22}).The various mass measurements are discussed in the ensuing subsections.

\begin{table}[t]
 \centering
 \caption{Mass measurements of M~31}\label{sample-table}
 {\tablefont\begin{tabular}{c|c|c|c|c}
    \midrule
    Publication & Method & d$\rm_{M~31}$ [kpc] & r$\rm_{enc}$ [kpc] & M$\rm_{tot}$ [10$^{12}$ M$\rm_{\odot}$]\\
    \midrule
    \citet{Babcock39} & RC -- HII regions & 210 & 22.8 & 0.102 \\
    \citet{Wyse42} & RC -- HII regions & 210 & 22.8 & 0.095 \\
    \citet{Sch54} & RC -- HII regions & 460 & 22.8 & 0.14 \\
    \citet{Lohmann54} & RC -- HII regions & 460 & 27.3 & 0.33 \\
    \citet{Schmidt57} & RC -- HI gas & 630 & 27.3 & 0.338 \\
    \citet{Brandt60} & RC -- HII regions \& HI gas & 600 & 34.2 & 0.37 \\
    \citet{Roberts66} & RC -- HI gas & 690 & 27.3 & 0.31 \\
    \citet{Rubin70} & RC -- HII regions & 690 & 29.6 & $0.185 \pm 0.01$ \\
    \citet{Einasto70} & RC -- HII regions \& HI gas & 692 & 34.2 & 0.2 \\
    \citet{Gott70} & RC -- HI gas & 690 & 34.2 & 0.22 \\
    \citet{Hartwick74} & GCs & 667 & 19.3 & $0.34 \pm 0.14$ \\
    \citet{Deharveng75} & RC -- HII regions & 690 & 22.8 & $0.163 \pm 0.015$ \\
    \citet{Roberts75} & RC -- HI gas & 690 & 34.2 & 0.37 \\
    \citet{Gunn75} & LG kinematics & 690 & -- & 4 \\
    \citet{Hodge75} & Satellites & 690 & -- & 6 \\
    \citet{Emerson76} & RC -- HI gas & 690 & 31.2 & 0.218 \\
    \citet{Rood79} & Satellites & 667 & -- & 0.338 \\
    \citet{Bahcall81} & Satellites & 667 & 100 & 1 \\
    \citet{vdB81} & GCs & 667 & 22.8 & $0.24 \pm 0.12$ \\
    & Satellites & 667 & -- & $0.75 \pm 0.39$ \\
    \citet{Braun91} & RC -- HI gas & 690 & 31.3 & $0.2 \pm 0.01$ \\
    \citet{Courteau99} & LG kinematics & 758 & -- & $1.33 \pm 0.18$ \\
    \citet{Evans00} & Satellites, GCs \& PNe & 770 & -- & $1.23^{+1.8}_{-0.6}$ \\
    \citet{Cote00} & Satellites & 780 & -- & $0.79 \pm 0.05$ \\
    \citet{Klypin02} & Mass model & 770 & -- & $1.6$ \\
    \citet{Evans03} & GCs & 770 & 100 & $1.2$ \\
     & Satellites & 770 & -- & $1.1$ \\
    \citet{Ibata04} & Substructure & 780 & 125 & $0.75^{+0.25}_{-0.13}$ \\
    \citet{Carignan06} & RC -- HI gas & 780 & 35 & $0.34$ \\
    \citet{Galleti06} & GCs & 784 & -- & $2.4$ \\
    \citet{Fardal06} & Substructure & 784 & 125 & $0.74 \pm 0.12$ \\
    \citet{Geehan06} & Mass model & 784 & -- & $0.71$ \\
    \citet{Seigar08} & Mass model & 784 & -- & $0.82$ \\
    \citet{Lee08} & GCs & 780 & 100 & $1.92^{+0.14}_{-0.13}$  \\
    \citet{Chemin09} & Mass model & 785 & -- & 1 \\
    \citet{Corbelli10} & Mass model & 785 & -- & 1.3 \\
    \citet{Watkins10} & Satellites & 785 & 300 & $1.4 \pm 0.4$ \\
    \citet{vdM12} & LG kinematics & 770 & -- & $1.54 \pm 0.39$ \\
    \citet{Tamm12} & Mass model & 785 & -- & $0.95 \pm 0.15$ \\
    \citet{Tol12} & Simulation & 785 & -- & $1.2^{+0.9}_{-0.7}$ \\
    \citet{Fardal13} & Substructure & 780 & -- & $1.99^{+0.52}_{-0.41}$ \\
    \citet{Vel13} & GCs & 780 & 200 & $1.35 \pm 0.35$ \\
    \citet{Diaz14} & LG kinematics & 780 & -- & $1.7 \pm 0.3$ \\
    \citet{Hayashi14} & Mass model & 785 & 200 & $1.82^{+0.49}_{-0.39}$ \\
    \citet{Pen14} & LG kinematics & 783 & -- & $1.5 \pm 0.3$ \\
    \citet{Sofue15} & RC & 770 & 31.3 & $1.99 \pm 0.39$ \\
    \citet{Pen16} & LG kinematics & 783 & -- & $1.33^{+0.39}_{-0.33}$ \\
    \citet{Kafle18} & PNe & 780 & -- & $0.8 \pm 0.1$ \\
    \citet{Zhai20} & Simulation & 787 & -- & $2.5^{+1.3}_{-1.1}$ \\
    \citet{VD21} & Simulation & -- & -- & $2.2^{+1.3}_{-0.8}$ \\
    \citet{Escala22} & Substructure & 785 & 36.6 & $0.926 \pm 0.278$ \\
    \citet{Patel22} & Simulation & 776 & -- & $3.02^{+1.3}_{-0.69}$ \\
    \citet{Dey23} & Substructure & 785 & 125 & $0.63^{+0.2}_{-0.13}$ \\
    \midrule
    \end{tabular}}
\tabnote{\textit{Notes}: Column 1: Source publication; Column 2: Method utilised for mass measurement; Column 3: Adopted distance to M~31 in source publication; Column 4: Radius within which enclosed mass has been measured, assuming present day best measured distance of 776~kpc \citep{Savino22}. The radius is not noted if the method, either directly or indirectly, measures the total mass of M~31.; Column 5: Mass of M~31, as reported.}
\label{table:mass}
\end{table}

\subsection{The disc rotation curve and early mass measurements of M~31}
\label{sect:early}

Following the first measurement of the M~31 disc RC by \citet{Babcock39}, the same data were modelled as a thin disc by \citet{Wyse42} leading to a mass estimate of $9.5 \times 10^{10}$~M$_{\odot}$ for M~31. These authors had assumed a variable mass-to-light ratio (M/L) as a function of radius in the M~31 disc. Assuming a constant M/L and thin disc for the M~31 RC from \citet{Babcock39} and additionally also from \citet{Mayall50}, \citet{Sch54} measured a mass of $1.4 \times 10^{11}$~M$_{\odot}$ for M~31. At that time, there was an ongoing tension in the measured distance of M~31, summarized in \citet{Hubble53}, and thus \citet{Sch54} had assumed a distance of 460~kpc for M~31, an average of the contentious distance estimates. With the same RC data and distance but with a slightly different disc mass model, \citet{Lohmann54} measured a mass of $3.3 \times 10^{11}$~M$_{\odot}$ for M~31. Soon after, \citet{vdH57} reported the first measurement of the HI gas RC of M~31 from their radio observations. The mass of M~31 based on the HI gas RC was reported by \citet{Schmidt57} to be  $3.38 \times 10^{11}$~M$_{\odot}$, similar to that of \citet{Lohmann54}. \citet{Brandt60} simultaneous modelled the M~31 disc RC as a thin disc from both HII regions \citep{Babcock39, Mayall50} and the HI gas \citep{vdH57} to obtain a similar mass of $3.7 \times 10^{11}$~M$_{\odot}$ for M~31. Both \citet{Schmidt57} and \citet{Brandt60} utilised the distance modulus measured for M~31 from cepheid variables by \citet{Baade55} but made additional different corrections for extinction (see Table~\ref{table:mass} for the values). \citet{Roberts66} obtained the HI gas RC out to $\sim$2~deg in M~31 and assuming a distance of 690 kpc \citep{Baade55} and similar model as \citet{Brandt60}, they obtained a mass of $3.1 \times 10^{11}$~M$_{\odot}$ for the galaxy.

\citet{Rubin70} utilised an efficient image-tube spectrograph to measure the radial velocities of individual HII regions in M~31 in reasonable time (1--1.5 hr per HII region). They constructed the RC of M~31 from these HII regions and determined a mass of $1.85 \pm 0.1 \times 10^{11}$~M$_{\odot}$ for M~31. While previous authors had generally termed their determined mass as the total (or near total) mass of M~31, \citet{Rubin70} had noted that their determined mass was the enclosed mass and that there should be more missing mass in the outskirts of the galaxy, which later became clearer as they and their contemporaries applied their efficient observation techniques to other galaxies \citep[see review by][]{Faber79}. However, \citet{Einasto70} simultaneously modelled the observed luminosity distribution of stars in M~31 and the RC from HII regions and HI gas to obtain a similar mass of $2 \times 10^{11}$~M$_{\odot}$ for M~31, but showing that no missing mass is required at least within the enclosed radius to reconcile the RC with the luminosity distribution of the galaxy. Similar values of the M~31 mass were computed by \citet{Gott70}, \citet{Roberts75} and \citet{Emerson76}, and later by \citet{Braun91} and even \citet{Carignan06} from their respective HI gas RC measurements, and by \citet{Deharveng75} for their RC constructed from HII regions (see Table~\ref{table:mass}). A distance of 690 kpc \citep{Baade55} to M~31 was generally accepted. These authors all quoted their mass obtained as being the enclosed mass out to their probed radius (see r$\rm_{enc}$ in Table~\ref{table:mass}).

At this time, bolstered by the suggestion of a massive dark halo required to support disc galaxies \citep{OP73}, alternative methods to measure the mass of galaxies started becoming popular. One of the earliest applied to M~31 was by \citet{Hartwick74} who applied the viral theorem to the globular cluster (GC) population of M~31 \citep[identified by][]{vdB69}. The GCs only enclosed a small radius in M~31 and they measured a similar mass as other authors did from the RCs (see Table~\ref{table:mass}). They had assumed a slightly lower distance of M~31 from \citet{Sandage71}. During this time, \citet{Gunn75} utilised the timing argument \citep{Kahn59} to simultaneously measure the mass of the MW and M~31, reporting a very high but uncertain mass of $4 \times 10^{12}$~M$_{\odot}$ for M~31. \citet{Hodge75} used the radial velocities of two dwarf satellites of M~31, NGC~247 and NGC~185, and applied tidal radius equations to obtain a similarly high and uncertain mass of $6 \times 10^{12}$~M$_{\odot}$ for M~31. \citet{Rood79} applied the virial theorem to the radial velocities of the six most luminous satellite galaxies of M~31  and mass of $3.38 \times 10^{11}$~M$_{\odot}$ for M~31, an uncertain value but one that is in-line with that measured in the central regions for M~31 from RCs. 

Such mass determination methods of M~31 and their development to present days are discussed individually in the ensuing subsections.

\subsection{Satellite radial velocity measurements}
\label{sect:sat}

Given that M~31 satellites occur at larger distances from the galaxy, they are thought to trace the mass of M~31 out to larger radii. The projected mass estimator method introduced by \citet{Bahcall81} brought an improvement to the viral mass technique used by \citet{Rood79}, determining the mass of M~31 as $\sim10^{12}$~M$_{\odot}$. The same projected mass estimator method was applied using an additional M~31 satellite by \citet{vdB81} to find slightly lower masses (see Table~\ref{table:mass}). Once more M~31 satellites were identified with measured radial velocities and the distance to M~31 was updated to 770~kpc (close to its present day value; \citealt{Holland98}), updated measurements of the mass of M~31 utilising variations of the projected mass estimator were reported (see Table~\ref{table:mass}) by \citet{Evans00} and \citet{Cote00}. Notably, \citet{Evans00} also utilised GCs and Planetary Nebulae (PNe) in M~31 to trace its mass. \citet{Evans03} developed the tracer mass estimator method and applied it to the satellites of M~31 to measure a mass of $1.1 \times 10^{12}$~M$_{\odot}$ for the M~31 galaxy. The latest iteration of this technique was applied by \citet{Watkins10} to obtain a mass of $\sim1.4 \times 10^{12}$~M$_{\odot}$ for M~31 out to ~300~kpc (well beyond their estimated virial radius of M~31).

\subsection{Globular cluster radial velocity measurements}
\label{sect:gc}

Similar to satellite galaxies, the projected mass estimator method \citep{Bahcall81} was applied to GCs to estimate the mass of M~31 enclosed by these tracers. \citet{vdB81} measured a mass of $2.4 \pm 0.12 \times 10^{11}$~M$_{\odot}$ for M~31 within $\sim22.8$~kpc, similar to that from RCs at this radius. \citet{Evans03} utilised the tracer mass estimator technique for GCs but now extrapolated out to ~100~kpc to find that M~31 has a mass of $1.2 \times 10^{12}$~M$_{\odot}$, consistent with that from applying the same technique to satellite galaxies. With an updated sample of GCs and slightly larger M~31 distance of 780~kpc \citep{Mcc05}, the same technique led to a larger measured mass of $2.4 \times 10^{12}$~M$_{\odot}$ for M~31 by \citet{Galleti06}, with a slightly lower value from \citet{Lee08}. The latest application of the tracer mass estimator technique to GCs by \citet{Vel13} fetched a slightly lower mass of $1.35 \pm 0.35 \times 10^{12}$~M$_{\odot}$ with the authors stating that the determined mass is highly dependent on the chosen model and assumptions within. 

\subsection{Mass models of M~31}
\label{sect:model}

Similar in spirit to the technique from \citet{Einasto70} but now including a bulge, an adiabatic disc and a cuspy halo component described by a NFW \citep{nfw96} halo profile, \citet{Klypin02} simultaneously fitted the surface brightness profile of M~31 and its RC measurements from CO and HI gas within $\sim$35~kpc. From their mass models, they thus obtained a mass of $1.6 \times 10^{12}$~M$_{\odot}$ for M~31. Similar models by \citet{Geehan06} and \citet{Seigar08}, however, led to lower modelled mass values of M~31 (see Table~\ref{table:mass}), nearly half that of \citet{Klypin02}. Applying similar mass models to their measured HI RCs out to $\sim40$~kpc from the center of M~31, both \citet{Chemin09} and \citet{Corbelli10} find slightly larger masses for M~31 of $10^{12}$~M$_{\odot}$ and $1.3 \times 10^{12}$~M$_{\odot}$ respectively. The value is consistent (see Table~\ref{table:mass}) with that of \citet{Tamm12} who utilised stellar population models to obtain a stellar mass map of M~31 using multi-wavelength photometry of M~31, and modelled the total mass of M~31 similar to \citet{Klypin02}. \citet{Hayashi14} modelled a prolate halo of M~31 to find a total mass of $\sim1.82 \times 10^{12}$~M$_{\odot}$ for the galaxy. They however noted that the choice of assumed halo density profile (NFW or otherwise) has strong bearing on the measured mass of M~31.

\subsection{Substructure radial velocity measurements}
\label{sect:subs}

From improvements in wide-field imaging techniques and instrumentation, a number of surveys conducted in the outskirts of M~31 led to the identification of a plethora of substructures in the halo of M~31 \citep[see][and references therein]{Mcc18}. The most prominent among these is the Giant Stream (GS). From their galaxy mass models, \citet{Ibata04} found that the kinematics of the GS require a total mass inside 125 kpc of $\sim7.5 \times 10^{11}$~M$_{\odot}$. Similar values were found by both \citet{Fardal06} and \citet{Dey23} using a similar technique and GS kinematics out to $\sim$125~kpc. Accounting for updated kinematic measurements of the GS out to larger radii and updated surface brightness measurements as constraints for mass models, \citet{Fardal13} found a virial mass of $\sim2 \times 10^{12}$~M$_{\odot}$ for M~31. The higher mass is also supported by the mass within $\sim37$~kpc determined by \citet{Escala22} from kinematics of the NE-Shelf substructure of M~31 (see Table~\ref{table:mass}).
 
\subsection{Local Group kinematics}
\label{sect:LG}
The relative motion of galaxies in the LG around its barycenter were utilised by \citet{Courteau99} to obtain the mass of the LG. They had assumed M~31 to be 1.5 times as massive as the MW and found that M~31 had a mass of $\sim1.3 \times 10^{12}$~M$_{\odot}$. With an estimate of the M~31 transverse and radial velocities towards the MW, \citet{vdM12} applied the timing argument \citep{Kahn59} to determine a mass of $\sim1.5 \times 10^{12}$~M$_{\odot}$ for M~31. A similar value was later determined by \citet{Pen14} with a slightly lower value when the LMC kinematics \citep{Pen16} are considered (see Table~\ref{table:mass}). By balancing mass and momentum in the LG, \citet{Diaz14} simultaneously determined the total mass of the LG as the mass ratio of the MW and M~31. They also found a similar mass of $\sim1.7 \times 10^{12}$~M$_{\odot}$ for M~31. 

\subsection{Cosmological simulations}
\label{sect:CS}
Over the past decade, a number of large scale cosmological simulations have developed, that are well-constrained to observed scaling relations for massive galaxies \citep[see review by][]{Vog20}. One of the earliest use of cosmological simulations to measure the mass of M~31 was from \citet{Tol12}, who actually used the virial theorem to determine the M~31 mass from satellite radial velocities but calculated a correction factor by applying the same technique to a simulated galaxy system and checking the discrepancy. Over time, techniques have developed to identify observed galaxy analogues in such cosmological simulations based on observed properties and thereby inferring unobserved properties from the simulated analogues. \citet{Zhai20} identified LG analogues in the Millenium simulations based on observed properties of the two galaxies and their mutual orbits. Their M~31 analogues thus identified with a radial in-fall into the MW had a mean mass of $\sim2.5 \times 10^{12}$~M$_{\odot}$ which the report as the mass of M~31. \citet{VD21} carry out a similar exercise but for the SIMBA and TNG simulations identifying the masses of M~31 analogues given the input positions of the observed M~31 and its three most massive satellites (M~33, M~32, M~110) as well as their radial velocities (for M~31, tangential velocities are also used as a constrain). A similar mean mass of $\sim2.2 \times 10^{12}$~M$_{\odot}$ is obtained for the analogues and attributed to M~31. \citet{Patel22} also use a similar approach but identify M~31 analogues in Illustris TNG-dark simulations constrained by the radial and tangential motions of four of its satellites -- M~33, NGC~185, NGC~147 \& IC~10. They find a mean mass of $\sim3 \times 10^{12}$~M$_{\odot}$ for the analogues and attribute it to M~31.

\subsection{Other techniques}
\label{sect:other}

\citet{Sofue15} combined the disc rotation velocities of M~31 with the radial velocities of its satellites and GCs to construct the grand rotation curve of M~31. They then used mass models for the disc, bulge, and an NFW halo to determine the mass of M~31 to be $\sim2 \times 10^{12}$~M$_{\odot}$. \citet{Kafle18} used high-velocity PNe to  measure the escape velocity of PNe as a function of radius, while also simultaneously constraining the virial mass and viral radius of M~31. They find a mass of $\sim8 \times 10^{11}$~M$_{\odot}$ for M~31. 

\begin{figure}[t]
  \centerline{\vbox to 1pc{\hbox to 1pc{}}}
  \includegraphics[width=\columnwidth]{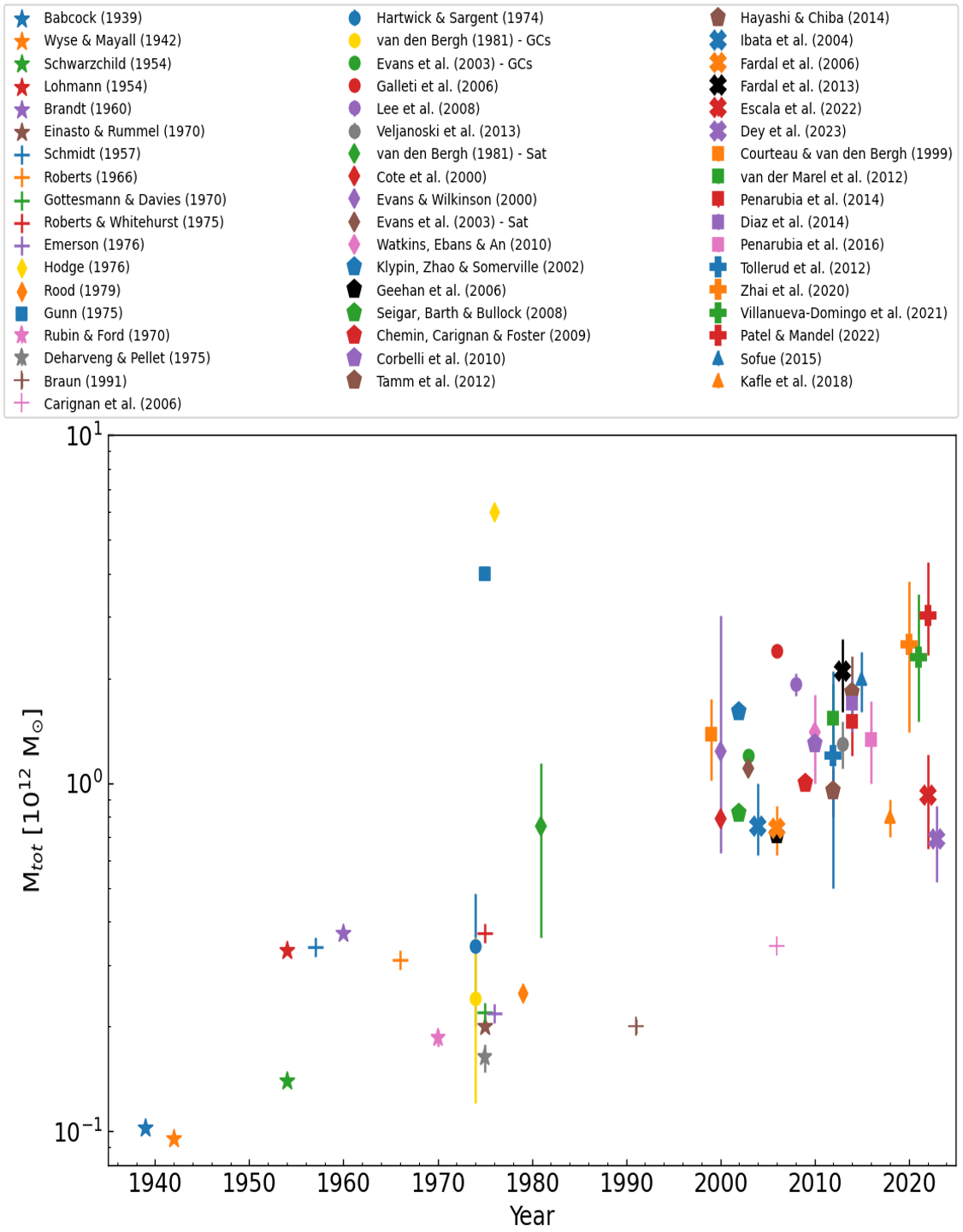}
  \caption{The measured mass of M~31 (in log scale) as a function of publication year. Different symbols refer to different mass measurement techniques that are described in Section~\ref{sect:mass}. Star: RC -- HII regions; Narrow plus: RC -- HI gas; Diamond: Satellites; Circle -- GCs ; Square -- LG kinematics; Pentagon -- Mass models; Cross -- Substructures; Broad plus -- simulations; Triangles -- others.}
  \label{fig:mass}
\end{figure}

\section{Discussion}
\label{sect:disc}
Figure~\ref{fig:mass} shows the reported mass of M~31 as a function of publication year. From the lowest measured mass of M~31 from \citet{Wyse42} to the highest value from \citet{Patel22}, the measured mass of M~31 has increased by almost ~30 times over the span of the past $\sim$80~years. The initial mass measurements were actually restricted to that of the luminous disc and inner regions of the galaxy. Probing greater and greater enclosed radius of M~31 (see Table~\ref{table:mass}) using the variety of techniques discussed in Section~\ref{sect:mass}, coupled with improvements in its measured distance, has resulted in a steady increase of the measured mass of M~31 (see Figure~\ref{fig:mass}). Most measurements of the M~31 total mass over the past $\sim$20~years do place its mass within 1--2~$\times 10^{12}$~M$_{\odot}$. 

\citet{Benisty22} find a LG mass of ${3.4}_{-1.1}^{+1.4} \times 10^{11}$~M$_{\odot}$ from LG analogues in the Illustris TNG simulations using particularly the M~31 tangential velocities as an important criteria in choosing said analogues. The same simulation suite is used by \citet{Patel22} and \citet{VD21} but using different conditions for choosing M~31 analogues to find M~31 masses of $\sim3$~\&~$\sim2.2$~$\times 10^{12}$~M$_{\odot}$ respectively, which are then nearly $\sim$88\%~\&$\sim$66\% of the mass of the LG. It is thus clear that the mass of M~31 determined using cosmological simulations depends heavily on the choice of priors. These large-scale cosmological simulations have their mean galaxy masses matched to the observed stellar-mass halo-mass relation. This observed relation would require a typical galaxy with the stellar mass of M~31 to have a halo mass of $\sim10^{13}$~M$_{\odot}$ \citep{McGaugh23}. Thus it is unsurprising that, despite including some priors, M~31 analogues in cosmological simulations tend to be on the more massive end (as the typical M~31 mass galaxies in these simulations lie in more massive haloes). Stronger priors may be necessary to obtain a more accurate measurement of the M~31 mass from their simulated analogues. This is especially true given that there is observational evidence that M~31 has undergone a recent major merger \citep{Bh19a,Bh19b,Bh21,Bh22,Bh23,Ar22} consistent with 2--3~Gyr old 1:4 major merger simulations \citep{ham18}. This is in sharp contrast to that of the MW \citep[see][for a summary]{Bh23}. Considering the merger history in choosing the simulated analogues of M~31 may yield more accurate mass measurements.

The most accurate total mass measurements of M~31 for the time being come from different techniques discussed in Section~\ref{sect:mass}. To reiterate, \citet{Watkins10} find a near-total mass of M~31 as 
$\sim1.4\times 10^{12}$~M$_{\odot}$ from applying the tracer mass estimator technique to the radial velocities of M~31 satellites (see Sect~\ref{sect:sat}). The same technique applied to GCs \citep{Vel13} also yields a similar M~31 mass (see Sect~\ref{sect:gc}). The mass models of M~31 constrained to the latest HI RC in the disc and the surface brightness profile also yield a similar mass (\citealt{Corbelli10};see Sect~\ref{sect:model}). A mass of $\sim2\times 10^{12}$~M$_{\odot}$ is determined from M~31 using models that are consistent with the GS kinematics (\citealt{Fardal13};see Sect~\ref{sect:subs}). From the timing argument, \citet{Pen16} find the mass of M~31 to be $\sim1.33\times 10^{12}$~M$_{\odot}$ (see Sect~\ref{sect:LG}). Finally, from the grand RC, \citet{Sofue15} measure a mass of $\sim2 \times 10^{12}$~M$_{\odot}$ for M~31. 

Considering the mean of the mass of M~31 determined from these works using the aforementioned different techniques, we find that M~31 has a mass of $\sim1.56 \times 10^{12}$~M$_{\odot}$. There is still a significant margin to improve the mass measurement of M~31 as understood from the previous discussion. 


\begin{thebibliography}{}
\bibitem[Arnaboldi \textit{et. al}(2022)]{Ar22}  Arnaboldi, M., Bhattacharya, S., Gerhard, O., \textit{et. al}\ 2022, {\it A\&A}, 666, A109
\bibitem[Baade \& Swope(1955)]{Baade55} Baade, W. \& Swope, H. H.\ 1955, \textit{AJ}, 60, 151
\bibitem[Babcock(1939)]{Babcock39} Babcock, H. W.\ 1939, Lick Observatory Bulletin, 498, 41
\bibitem[Bahcall \& Tremaine(1981)]{Bahcall81} Bahcall, J. N. \& Tremaine, S.\ 1981, \textit{ApJ}, 244, 805
\bibitem[Benisty \textit{et. al}(2022)]{Benisty22} Benisty, D., Vasiliev, E., Evans, N. W., \textit{et. al}\ 2022, {\it ApJ}, 928, L5
\bibitem[Bhattacharya \textit{et. al}(2019a)]{Bh19a} Bhattacharya, S., Arnaboldi, M., Hartke, J., \textit{et. al}\ 2019, {\it A\&A}, 624, A132
\bibitem[Bhattacharya \textit{et. al}(2019b)]{Bh19b} Bhattacharya, S., Arnaboldi, M., Caldwell, N., \textit{et. al}\ 2019, {\it A\&A}, 631, A56
\bibitem[Bhattacharya \textit{et. al}(2021)]{Bh21} Bhattacharya, S., Arnaboldi, M., Gerhard, O., \textit{et. al}\ 2021, {\it A\&A}, 647, A130
\bibitem[Bhattacharya \textit{et. al}(2022)]{Bh22} Bhattacharya, S., Arnaboldi, M., Caldwell, N., \textit{et. al}\ 2022, {\it MNRAS}, 517, 2343
\bibitem[Bhattacharya \textit{et. al}(2023)]{Bh23} Bhattacharya, S., Arnaboldi, M., Hammer, F., \textit{et. al}\ 2023, arXiv:2304.14151
\bibitem[Brandt(1960)]{Brandt60} Brandt, J. C.\ 1960,\textit{ AJ}, 131, 293
\bibitem[Braun(1991)]{Braun91} Braun, R.\ 1991, \textit{ApJ}, 372, 54
\bibitem[Carignan~{\it et. al}(2006)]{Carignan06} Carignan, C., Chemin, L., Huchtmeier, W. K., {\it et. al}\ 2006, \textit{ApJ}, 641, L109
\bibitem[Corbelli~{\it et. al}(2010)]{Corbelli10} Corbelli, E., Lorenzoni, S., Walterbos, R., {\it et. al}\ 2010, \textit{A\&A}, 511, A89
\bibitem[Cote~{\it et. al}(2000)]{Cote00} Cote, P., Mateo, M., Sargent, W. L. W., {\it et. al}\ 2000, \textit{ApJ}, 537, 91
\bibitem[Courteau \& van den Bergh(1999)]{Courteau99} Courteau, S. \& van den Bergh, S.\ 1999, \textit{AJ}, 118, 337
\bibitem[Chemin, Carignan \& Foster(2009)]{Chemin09} Chemin, L., Carignan, C. \& Foster, T.\ 2009, \textit{ApJ}, 705, 1395
\bibitem[Deharving \& Pellet(1975)]{Deharveng75} Deharveng, J. \& Pellet, A.\ 1975, \textit{A\&A}, 38, 15
\bibitem[Dey(2023)]{Dey23} Dey A., Najita, J. R., Koposov, S. E., {\it et. al}\ 2023,\textit{ApJ}, 944, 1
\bibitem[Diaz~{\it et. al}(2014)]{Diaz14} Diaz, J. D., Koposov, S. E., Irwin, M., {\it et. al}\ 2014, \textit{MNRAS}, 443, 1688
\bibitem[Einasto \& Rummel(1970)]{Einasto70} Einasto J., \& Rummel U.\ 1970, in Becker W., Kontopoulos G. I., eds, Proc.
IAU Symp. 38, The Spiral Structure of Our Galaxy. Reidel, Dordrecht, p. 51
\bibitem[Emerson(1976)]{Emerson76} Emerson, D. T.\ 1976, \textit{MNRAS}, 176, 321
\bibitem[Escala~{\it et. al}(2022)]{Escala22} Escala, I., Gilbert, K. M., Fardal, M. A., {\it et. al}\ 2022, \textit{AJ}, 164, 20
\bibitem[Evans \& Wilkinson(2000)]{Evans00} Evans, N. W. \& Wilkinson, M. I.\ 2000, \textit{MNRAS}, 316, 929
\bibitem[Evans~{\it et. al}(2003)]{Evans03} Evans, N. W., Wilkinson, M. I., Perrett, K. M., {\it et. al}\ 2003, \textit{ApJ}, 583, 752
\bibitem[Faber \& Gallagher(1979)]{Faber79} Faber, S. M. \& Gallagher, J. S.\ 1979, \textit{AR\&AA}, 17, 135
\bibitem[Fardal~{\it et. al}(2006)]{Fardal06} Fardal, M. A., Babul, A., Geehan, J. J. {\it et. al}\ 2006, \textit{MNRAS}, 366, 1012
\bibitem[Fardal~{\it et. al}(2013)]{Fardal13} Fardal, M. A., Weinberg, M. D., Babul, A., {\it et. al}\ 2013, \textit{MNRAS}, 434, 2779
\bibitem[Galleti~{\it et. al}(2006)]{Galleti06} Galleti, S., Federici, L., Bellazzini, M. {\it et. al}\ 2006, \textit{A\&A}, 456, 985
\bibitem[Geehan~{\it et. al}(2006)]{Geehan06} Geehan, J. J., Fardal, M. A., Babul, A., {\it et. al}\ 2006, \textit{MNRAS}, 366, 996
\bibitem[Gottesman \& Davies(1970)]{Gott70} Gottesman, S. \& Davies, R.\ 1970, \textit{MNRAS}, 149, 263
\bibitem[Gunn(1975)]{Gunn75} Gunn, J.\ 1975, Comments Ap. Sp. Phys., 6, 7
\bibitem[Hammer(2018)]{ham18} Hammer F., Yang Y. B., Wang J. L., \textit{et. al}\ 2018, \textit{MNRAS}, 475, 2754
\bibitem[Hartwick \& Sargent(1974)]{Hartwick74} Hartwick, F. \& Sargent, W.\ 1974, \textit{ApJ}, 190, 283
\bibitem[Hayashi \& Chiba(2014)]{Hayashi14} Hayashi, K. \& Chiba, M.\ 2014, \textit{ApJ}, 789, 62
\bibitem[Hodge(1975)]{Hodge75} Hodge, P. W.\ 1975, Bulletin of the American Astronomical Society, 7, 506
\bibitem[Hodge(1992)]{Hodge92} Hodge P. W., ed.\ 1992, Astrophysics and Space Science Library Vol. 176. Kluwer, Dordrecht
\bibitem[Holland(1998)]{Holland98} Holland, S.\ 1998, \textit{AJ}, 115, 1916
\bibitem[Hubble(1925)]{Hubble25} Hubble, E. P.\ 1925, The Observatory, 48, 139
\bibitem[Hubble(1929)]{Hubble29} Hubble, E. P.\ 1929, \textit{ApJ}, 69, 103
\bibitem[Hubble \& Sandage(1953)]{Hubble53} Hubble, E. P. \& Sandage, A.\ 1953, \textit{ApJ}, 118, 353
\bibitem[Ibata~{\it et. al}(2004)]{Ibata04} Ibata, R., Chapman, M., Ferguson, A. M. N., {\it et. al}\ 2004, \textit{MNRAS}, 351, 117
\bibitem[Kafle~{\it et. al}(2018)]{Kafle18} Kafle, P. R., Sharma, S., Lewis, G. F., {\it et. al}\ 2018, \textit{MNRAS}, 475, 4043
\bibitem[Kahn \& Woltjer(1959)]{Kahn59} Kahn, F. D. \& Woltjer, L.\ 1959, \textit{ApJ}, 130, 705
\bibitem[Klypin, Zhao \& Somerville(2002)]{Klypin02} Klypin, A., Zhao H. \& Somerville R. S.\ 2010, \textit{ApJ}, 573, 597
\bibitem[Lee~{\it et. al}(2008)]{Lee08} Lee, M. G., Hwang, H. S., Kim, S. C. {\it et. al}\ 2008, \textit{ApJ}, 674, 886
\bibitem[Lohmann(1954)]{Lohmann54} Lohmann, W.\ 1954, Zeitschrift für Astrophysik, 35, 159
\bibitem[Mateo(1998)]{Mateo98} Mateo, M.\ 1998, \textit{ARA\&A}, 36, 435
\bibitem[Mayall(1950)]{Mayall50} Mayall, N. U.\ 1950, Publication of the Observatory of the University of Michigan, 10, 19
\bibitem[McConnachie~{\it et. al}(2005)]{Mcc05} McConnachie, A. W., Irwin, M. J., Ferguson, A. M. N., {\it et. al}\ 2005, \textit{MNRAS}, 356, 979
\bibitem[McConnachie~{\it et. al}(2018)]{Mcc18} McConnachie, A. W., Ibata, R., Martin, N., \textit{et. al}\ 2018, {\it ApJ}, 868, 55
\bibitem[McGaugh(2023)]{McGaugh23} McGaugh, S.\ 2023, arXiv:2305.00858
\bibitem[Navarro, Frenk \& White(1996)]{nfw96} Navarro, J. F., Frenk, C. S., \& White, S. D. M.\ 1996, \textit{ApJ}, 462, 563
\bibitem[Ostriker \& Peebles(1973)]{OP73} Ostriker, J. P., \& Peebles, P. J. E. \ 1973, \textit{ApJ}, 186, 467
\bibitem[Patel \& Mandel(2022)]{Patel22} Patel, E. \& Mandel, K. S.\ 2022, arXiv:2211.15928
\bibitem[Penarrubia~{\it et. al}(2014)]{Pen14} Penarrubia, J., Ma, Y. Z., Walker, M. G., {\it et. al}\ 2014, \textit{MNRAS}, 443, 2204
\bibitem[Penarrubia~{\it et. al}(2016)]{Pen16} Penarrubia, J., Gomez, F. A., Besla, G., {\it et. al}\ 2016, \textit{MNRAS}, 456, L54
\bibitem[Roberts(1966)]{Roberts66} Roberts, M. S.\ 1966, \textit{ApJ}, 144, 639
\bibitem[Roberts \& Whitehurst(1975)]{Roberts75} Roberts, M. \& Whitehurst, R.\ 1975, \textit{ApJ}, 201, 327
\bibitem[Rood(1979)]{Rood79} Rood, H.\ 1979, \textit{ApJ}, 232, 699
\bibitem[Rubin \& Ford(1970)]{Rubin70} Rubin, V. \& Ford, W.\ 1970, \textit{ApJ}, 159, 379
\bibitem[Sandage \& Tammann(1971)]{Sandage71} Sandage, A. \& Tammann, G. A.\ 1971, \textit{ApJ}, 167, 293
\bibitem[Savino(2022)]{Savino22} Savino, A., Weisz, D. R., Skillman, E. D., {\it et. al}\ 2022, \textit{ApJ}, 938, 101
\bibitem[Schwarzschild(1954)]{Sch54} Schwarzschild, M.\ 1954,\textit{ AJ}, 59, 273
\bibitem[Schmidt(1957)]{Schmidt57} Schmidt, M.\ 1957, Bulletin of the Astronomical Institutes of the Netherlands, 14, 17
\bibitem[Seigar, Barth, \& Bullock(2008)]{Seigar08} Seigar, M. S., Barth, A. J., \& Bullock, J. S.\ 2008, \textit{MNRAS}, 389, 1911
\bibitem[Sofue(2015)]{Sofue15} Sofue, Y.\ 2015,\textit{PASJ}, 67, 65
\bibitem[Tamm~{\it et. al}(2012)]{Tamm12} Tamm, A., Tempel, E., Tenjes, P., {\it et. al}\ 2012, \textit{A\&A}, 536, A4
\bibitem[Tollerud~{\it et. al}(2012)]{Tol12} Tollerud, E. J., Beaton, R. L., Geha, M. C., {\it et. al}\ 2012, \textit{ApJ}, 752, 45
\bibitem[van de Hulst, Raimond \& van Woerden(1957)]{vdH57} van de Hulst, H. C., Raimond, E. \& van Woerden, H.\ 1957, Bulletin of the Astronomical Institutes of the Netherlands, 14, 1
\bibitem[van den Bergh (1969)]{vdB69} van den Bergh, S.\ 1969, \textit{ApJS}, 19, 145
\bibitem[van den Bergh (1981)]{vdB81} van den Bergh, S.\ 1981, \textit{PASP}, 93, 428
\bibitem[van der Marel~{\it et. al}(2012)]{vdM12} van der Marel, R. P., Fardal, M., Besla, G., {\it et. al}\ 2012, \textit{ApJ}, 753, 14
\bibitem[Veljanoski~{\it et. al}(2013)]{Vel13} Veljanoski, J.,  Ferguson, A. M. N., Mackey, A. D., {\it et. al}\ 2013, \textit{ApJL}, 768, 33
\bibitem[Villanueva-Domingo~{\it et. al}(2021)]{VD21} Villanueva-Domingo, P.,  Villaescusa-Navarro, F., Genel, S., {\it et. al}\ 2021, arXiv:2111.14874
\bibitem[Vogelsberger~{\it et. al}(2020)]{Vog20} Vogelsberger, M., Marinacci, F., Torrey, P., {\it et. al}\ 2020, Nature Reviews Physics, 2, 42
\bibitem[Watkins, Evans \& An(2010)]{Watkins10} Watkins, L. L., Evans, N. W. \& An, J. H.\ 2010, \textit{MNRAS}, 406, 264
\bibitem[Wyse \& Mayall(1942)]{Wyse42} Wyse, A. B. \& Mayall, N. U.\ 1942, \textit{ApJ}, 95, 24
\bibitem[Zhai~{\it et. al}(2020)]{Zhai20} Zhai, M.,  Guo, Q., Zhao, G., {\it et. al}\ 2020, \textit{ApJ}, 890, 27
\end{thebibliography}
\end{document}